\documentclass[doublecol]{epl2} 
\usepackage{latexsym,amssymb,amsfonts,color,graphicx,amsmath}
\usepackage{upgreek}

\title{Dynamics of squirmers in explicitly modeled polymeric fluids}
\shorttitle{Dynamics of squirmers in explicitly modeled polymeric fluids} 

\author{A. Z\"ottl\inst{1,2}}
\shortauthor{A. Z\"ottl}

\institute{                    
  \inst{1} Faculty of Physics, University of Vienna, Kolingasse 14-16, 1090 Wien, Austria \\
  \inst{2} Institute for Theoretical Physics, TU Wien, Wiedner Hauptstr.~8-10, 1040 Wien, Austria
}

\abstract{
  Biological microswimmers such as bacteria and sperm cells often encounter complex biological fluid environments.
  Here we use the well-known squirmer microswimmer model to show the importance of the local fluid microstructure and non-continuum effects on their swimming speed in different polymeric and filamentous fluids.
  Surprisingly, we find that different squirmer types move at considerably different speed in filamentous fluids which cannot be explained by existing continuum models, but by considering the local fluid and polymer properties around the squirmers. Furthermore, direct squirmer-polymer interactions slow down in particular pushers by trapping large stiff filaments in a self-generated recirculation region in front of them.
}

\begin{document}

\maketitle

\section{Introduction}
Many microorganisms swim in viscous fluids,
and their dynamics can often be described by low Reynolds number hydrodynamics \cite{Lauga2009a,Elgeti2015b}.
Furthermore, active colloids and droplets have been investigated extensivly as well-controllable model systems to study single  and collective microswimmer behavior  \cite{Zottl2016,Bechinger2016}.
A common minimal model to capture the hydrodynamics of biological and artificial microswimmers is the so-called squirmer  \cite{Lighthill1952,Blake1971a,Pedley2016}.
By applying different surface velocity fields to a rigid sphere,
different types of microswimmers can be modeled: pushers such as  bacteria, pullers such as algae, or source-dipole swimmers as leading order models for active colloids \cite{Zottl2016}.
For a squirmer moving in bulk analytic solutions for the swimming speed and flow field exist,
and the squirmer has so far extensivly been employed to capture effects of self-generated hydrodynamic flow fields on microswimmer dynamics \cite{Zottl2016,Pedley2016}.

Often microswimmers move through more complex fluids which sometimes fail to be described as a simple continuous Newtonian fluid.
Recently there has been great interest in understanding the behavior and dynamics of microswimmers in such fluids, which is less well understood compared to swimming in Newtonian fluids \cite{Li2021b}.
So far the squirmer moving in complex, viscoelastic, and non-homogeneous fluids has been studied theoretically and by using continuum fluid models.
Examples for non-homogeneous but Newtonian fluids include a squirmer moving in  concentric viscous fluid layers \cite{Reigh2017b}, or  in viscosity gradients \cite{Datt2019,Eastham2020}.
Furthermore viscoelastic  \cite{Zhu2011,Zhu2012,DeCorato2015} and  shear-thinning effects \cite{Montenegro-Johnson2013,Datt2015xx,Ouyang2018}
have been investigated, as well as porosity modeled by Brinkman theory \cite{Leshansky2009,Nganguia2018,Nganguia2020}.

Polymeric fluids are often inhomogeneous and consist of nano- and microstructures as a result of the specific polymer composition. In particular in complex biological fluids such as mucus  or in  collagen fiber solutions  the size of these structures
  can be strongly inhomogeneous and be up to some micron  \cite{Kirch2012,Mickel2008}.

Furthermore, in  out-of-equlibrium situations such as for driven or active particles
moving in supramolecular solutions, the polymer relaxation time can be larger compared to the typical time a driven or active particle spends to move its own size,
which can leave behind a polymer-free zone \cite{Gutsche2008,Zottl2019b,Zottl2019}.
To capture heterogeneity of biological fluids such as mucus, two-fluid models have been employed which capture density- and viscosity-inhomogeneities around squirmers \cite{Reigh2017b}, as well as  porosity \cite{Nganguia2020a}.
  However, such models so far do not capture non-isotropic effects induced by the aforementioned wake depletion of polymers, which depends on the specific local flow field created by different squirmer types.

To capture non-continuum effects, we conduct in this work hydrodynamic
  simulations of explicitely modeled polymers around moving squirmers.
  We use a coarse-grained particle-based method called multiparticle collision dynamics (MPCD) \cite{Malevanets1999} where polymers can be coupled to a background fluid very efficiently in the presence of thermal noise \cite{Malevanets2000,Gompper2009}.
MPCD has been used in the last years extensively to model the hydrodynamics of microswimmers \cite{Zottl2020b}.
Important examples are bacteria \cite{Hu2015}, sperm \cite{Elgeti2010}, pathogens \cite{Babu2012}, Janus particles \cite{Huang2016} and squirmers \cite{Zottl2018}.
For example, squirmers in MPCD fluids have been succesfully modeled to study swimming and  persistent random walks \cite{Downton2009a,Goetze2010}, pairwise interactions \cite{Goetze2010}, dynamics in channel flow \cite{Zottl2012}, near surfaces \cite{Schaar2015}, under gravity \cite{Ruhle2018}, or in nematic fluids \cite{Mandal2021a}, as well as their
collective motion \cite{Zottl2014,Blaschke2016,Theers2018,Zantop2021a,Qi2022}.
MPCD also allows to model out-of-equilibrium multi-component systems, such as active or driven particles in polymer solutions \cite{Zottl2019,Zottl2019b,Qi2020}.
Recently the orientational dynamics of squirmers in solutions of polymers has been studied \cite{Qi2020}, motivated by experiments of active colloids in viscoelastic polymer solutions \cite{Gomez-Solano2016a}.

Here we investigate the dynamics of squirmers in explicitly modeled   fluids
including coarse-grained polymers consisting of a relatively small number of relatively large monomers (see Fig.~\ref{Fig:sys})  where entanglements and viscoelastic effects are expected to play a minor role, but non-homogeneous effects are present, depending on polymer density and stiffness in different self-generated squirmer fluid flows. 
We use the local fluid and polymer properties measured in the vicinity of the squirmer to determine underlying physical mechanisms for the observed swimming behavior.
We find considerable differences in swimming speeds between pushers and pullers, which can not be explained by existing continuum fluid models.
We show that local viscosity gradients slow down squirmers, as well as trapping and steric hindrance of large stiff polymers in the recirculation region in front of pushers.

\section{Model}
\begin{figure}
  \includegraphics[width=\columnwidth]{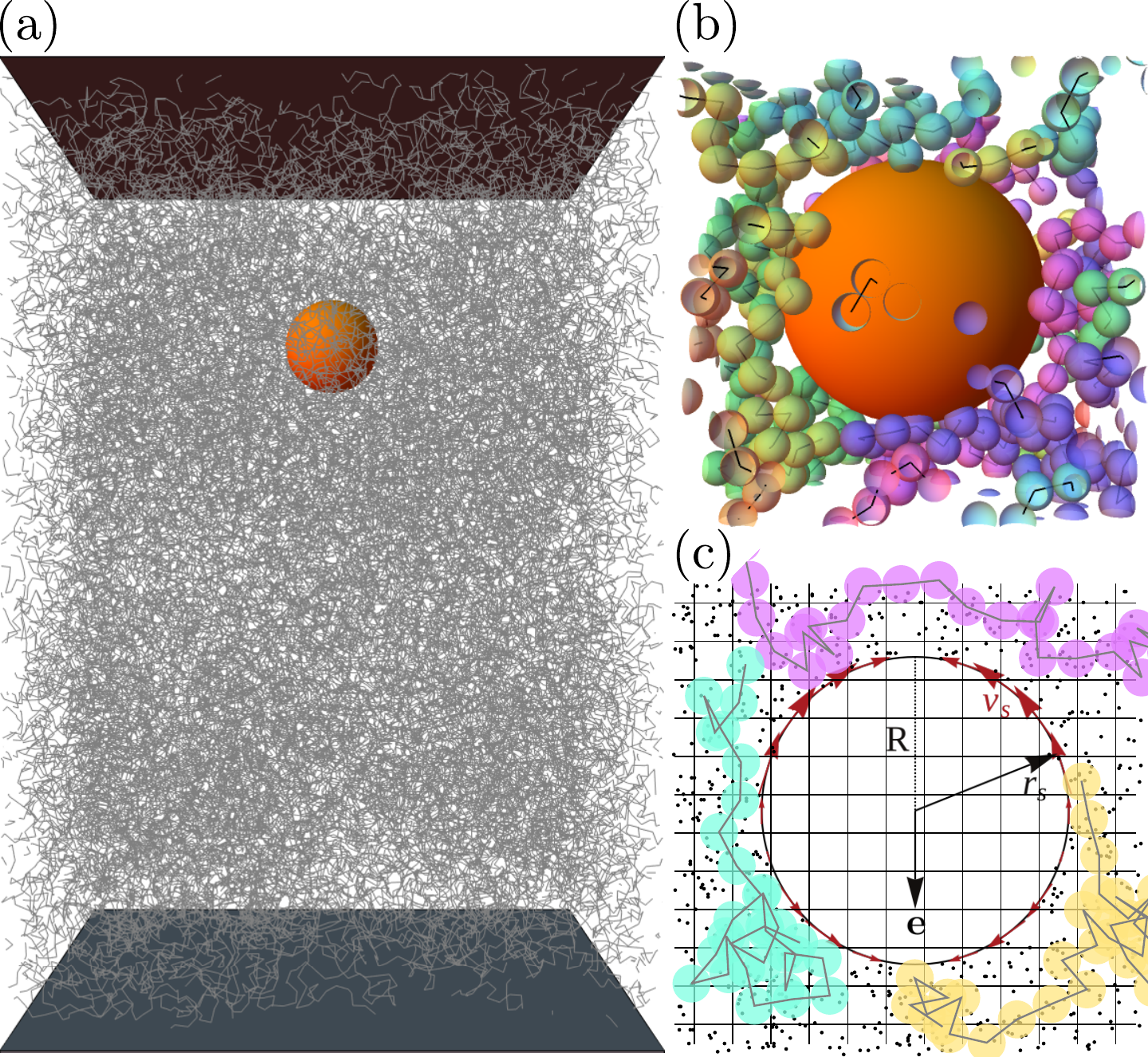}
  \caption{(a) Motion of a squirmer (orange) in the simulation box bounded by two rigid walls at $x=\pm S_X/2$ and filled with a MPCD fluid (not shown) and polymers (gray polygons).
    (b) Typical local environement around a squirmer. Individual bead-spring polymers are shown in different colors.
    (c) 2D sketch of the multi-component system: squirmer oriented along direction $\mathbf{e}$ with surface velocity $\mathbf{v}_s$ at surface $\mathbf{r}_s$, polymers (yellow, purple, cyan), and MPCD fluid particles (black dots). The grid for the collision step is shown in black. Not shown are virtual particles inside the squirmer used in the collision step.}
  \label{Fig:sys}
\end{figure}

We consider a coupled multi-component system, consisting of (i) a rigid squirmer, (ii) bead-spring polymers and filaments, (iii) the MPCD background fluid, and (iv) two bounding walls located at $x=\pm S_X/2$, where $\mathcal{V}=S_XS_YS_Z$ is the total volume of the simulation box with periodic boundary conditions in $y$- and $z$-direction, see Fig.~\ref{Fig:sys}.
The simulation box size is fixed by $S_X=72a_0$, $S_Y=S_Z=48a_0$.
The two walls are included to avoid a possible net fluid-flow and an associated violation of angular momentum conservation along the $x$ direction, i.e.\ the average squirmer direction \cite{Zottl2019b}.
In the following we desribe how we model the individual components and how they are coupled through hydrodynamic and steric interactions.

\subsection{MPCD}
Squirmer and polymers are immersed in a Newtonian background fluid such as water, which is modeled by MPCD and solves the Navier Stokes equations on a coarse-grained level including thermal fluctuations \cite{Malevanets2000,Kapral2008,Gompper2009}.
The fluid is represented by $N_f$ effective, pointlike fluid particles with mass $m$, positions $\mathbf{x}_i$ and velocities $\mathbf{v}_i$ with $i=1,\dots, N_f$.
The basic dynamics of the fluid particles consists of alternating streaming and collision steps. In the streaming step they move ballistically for a time $\delta t$ and their positions are updated according to
\begin{equation}
  \mathbf{x}_i (t + \delta t) = \mathbf{x}_i (t) + \mathbf{v}_i(t) \delta t \, .
  \label{Eq:str}
\end{equation}
After the streaming step particles are sorted into cubic cells of length $a_0$ (see Fig.~\ref{Fig:sys}(c)), and in the collision step all particles in a cell exchange momentum which updates their velocities
\begin{equation}
\mathbf{v}_i (t + \delta t) = \mathbf{v}_{\xi} (t) + \mathbf{v}_r(t) + \mathbf{v}_P(t) + \mathbf{v}_L(t),
  \label{Eq:col}
\end{equation}
where $\mathbf{v}_{\xi}=(1/N_{\xi})\sum_{j}\mathbf{v}_j$ is the instantaneous average velocity in the cell with $N_\xi$ the total number of fluid particles in a cell, and $j=1,\dots,N_{\xi}$ considers all particles in cell $\xi$.
The temperature $T$ in the cell is kept constant  using an Anderson thermostat, i.e.\ using random velocities $\mathbf{v}_r$ drawn from a Maxwell-Boltzmann distribution   with  variance $k_BT/m$.
The terms $\mathbf{v}_P $ and  $\mathbf{v}_L $ ensure local linear and angular momentum conservation, respectively.
In order to restore Galilean invariance and to minimize correlation effects we perform a random shift of the cell grid \cite{Gompper2009}.
Details of the algorithm are given in the SI.

As basic units of length, mass and energy we choose $a_0$, $m$ and $k_BT$, respectively, and times are measured in units of $t_0=\sqrt{ m a_0^2 /k_B T}$.
The free fluid model parameters are then the streaming time step  $\delta t$ and the average number of particles in a cell $n=\langle N_\xi \rangle$.
We use $\delta t = 0.02 t_0$ and  $n = 10$ which models viscous flows at  low Reynolds number
$\textnormal{Re}$
and sufficiently high Schmidt number $\textnormal{Sc}$ \cite{Padding2006,Gompper2009}.
In the absence of polymers the viscosity is then $\eta_0=16.04\sqrt{mk_BT/a_0^4}$ \cite{Zottl2014b} and $\textnormal{Sc}\approx 120$ \cite{Noguchi2008}.

\subsection{Polymer model}
We model polymers and filaments as $N$ beads of diameter $\sigma=a_0$ and mass $m_p=10m$ connected by stiff harmonic springs with rest length $l_0=\sigma$ and spring constant $k_{bond}=10^5k_BT/a_0^2$,
\begin{equation}
  V_{bond} = \frac 1 2 k_{bond}\sum_{i=2}^{N}(|\Delta \mathbf{r}_i|-l_0)^2
  \label{Eq:Bond}
\end{equation}
where $\mathbf{r}_i$, $i=1,\dots,N$ are the bead positions and  $\Delta \mathbf{r}_i = \mathbf{r}_i - \mathbf{r}_{i-1}$.
The total number $N_p$ of polymers is varied to set the desired polymer  density $\rho=N N_p \pi \sigma^3/(6(\mathcal{V}-\mathcal{V}_{Sq}))$, defined as the volume fraction of all monomers in the simulation box with $\mathcal{V}_{Sq}$ the squirmer volume.
To model
semiflexible polymers and filaments we use a bending potential
\begin{equation}
V_{b} = \frac 1 2 k_b\sum_{i=3}^{N}\left(\frac{\Delta \mathbf{r}_i \cdot \Delta \mathbf{r}_{i-1}}{|\Delta \mathbf{r}_i| |\Delta \mathbf{r}_{i-1}|} -1\right)^2
\label{Eq:BendP}
\end{equation}
where $k_b$ is the bending stiffness.
In total we consider six different  polymer models, i.e.\ flexible polymers ($k_b=0$) of length $N=\{12, 30, 100 \}$, and semiflexible filaments of length $N=30$ with stiffness $k_b=\{ 30k_BT, 300k_BT, 3000k_BT \}$.
Polymers are considered at different densities $\rho=\{0.05, 0.1,0.2,0.3 \}$.

All polymer beads interact with each other through a purely repulsive soft Weeks-Chandler-Anderson (WCA) potential \cite{Weeks1971},
\begin{equation}
 V_{WCA}(r) = 
 4\epsilon_0 \left[ {\left( \frac{\sigma^\ast}{r} \right)}^{12} - {\left( \frac{\sigma^\ast}{r}\right)}^{6} \right]  + \epsilon_0
\label{Eq:WCA}
\end{equation}
for distances between beads $r < \sigma$ and zero otherwise.
We use
$\epsilon_0=k_BT$ and $\sigma^\ast=\sigma/2^{1/6}$.

\subsection{Squirmer model}
We employ the simplest form of a squirmer, a rigid sphere of radius $R=4a_0$ with  orientation $\mathbf{e}$ and a static, axisymmetric and tangential surface velocity $\mathbf{v}_s=B_1(1+\beta \mathbf{e}\cdot \hat{\mathbf{r}}_s)[(\mathbf{e}\cdot \hat{\mathbf{r}}_s)\hat{\mathbf{r}}_s-\mathbf{e}]$ (see also Fig.~\ref{Fig:sys}(c)) which pushes fluid backwards and moves the squirmer forwards,
and depends on parameters $B_1>0$ and $\beta$.
In bulk continuum fluids of viscosity $\eta$ at $\textnormal{Re}=0$ the squirmer swims along its direction $\mathbf{e}$ with speed $V_b=\frac 2 3 B_1$, which is independent of  $\eta$ and  $\beta$.
However, $\beta$ determines the type of the flow field, and the squirmer is called a \textit{pusher} for $\beta<0$, a \textit{puller} for $\beta>0$, and a neutral squirmer for $\beta=0$ \cite{Zottl2016,Zottl2018,Zottl2023}.
The flow field of a squirmer is independent of viscosity $\eta$, but the power consumption for a fixed $\mathbf{v}_s$ to maintain the flow field  depends linearly on $\eta$ \cite{Downton2009a}.
For a pusher the far-field flow is that of an extensile force dipole, for a puller a contractile force dipole, and for a neutral squirmer a source dipole  \cite{Zottl2016}.
We set the squirmer parameters $B_1=0.062a_0/t_0$ and  $\beta=\{-3,0,3 \}$.
In the absence of polymers  the  Reynolds number is $\textnormal{Re}_0=V_bR\rho_f / \eta
  \approx 0.10$ with $\rho_f=m_0n/a_0^3$, and $\textnormal{Re} < \textnormal{Re}_0$ in the presence of polymers.

\begin{figure}
  \includegraphics[width=\columnwidth]{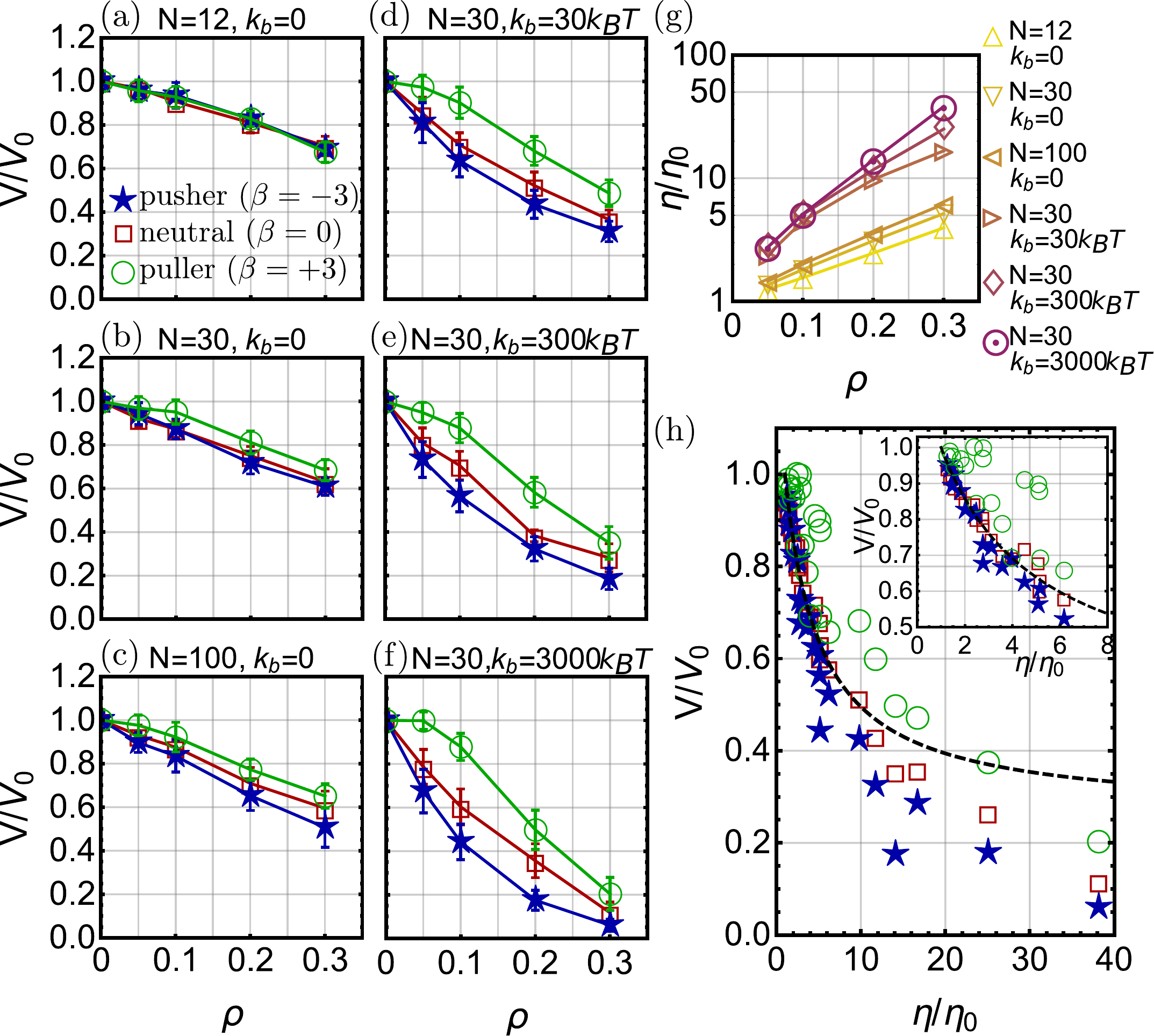}
  \caption{(a-f) Mean swimming speed $V$ of neutral squirmers ($\beta=0$), pushers ($\beta=-3$) and pullers ($\beta=3$) in  fluids of different length $N$ and bending stiffness $k_b$ depending on polymer density $\rho$, normalized by the swimming speed $V_0$ in the absence of polymers. Error bars indicate the standard deviations of time-averaged steady-state swimming speeds of 16 independent realizations. (g) Measured fluid viscosities $\eta$ normalized by the polymer-free viscosity $\eta_0$. (h) Swimming speed $V$ depending on viscosity $\eta$.}
  \label{Fig:vf}
\end{figure}

\subsection{Simulation procedure}
Initially the squirmer is located at $\mathbf{R}_0=\{ 18a_0,0,0 \}$ pointing in negative x direction ($\mathbf{e}_0=\{-1,0,0\}$).
Before the actual simulation we initialize and equlibrate for each parameter set an ensemble of 16 independent polymer solutions: Fluid particles and polymers are randomly 
distributed in the simulation box,  but are not allowed to overlap with the squirmer.
We then pre-equilibrate polymers for at least $5\cdot 10^4$ MPCD time steps; here we replace the squirmer by a hard non-moving sphere of radius $R$.
After equilibration we perform the actual simulations for $N_t=5\cdot 10^4$ MPCD time steps.
The typical squirmer  persistence time  \cite{Zottl2016}  $\tau_r =4\pi R^3 \eta /(k_BT) \approx 1.3 \cdot 10^4 t_0$ is much larger than the  simulation time $\Delta T = N_t\delta t=1000t_0$, and the associated P\'eclet number is $\textnormal{Pe}=3V_b\tau_r/(2R)\approx 200$  \cite{Zottl2016}. 
  Thus the squirmer   moves more or less along the $-x$ direction for a distance $\Delta x \lesssim V_b \Delta T \approx 40a_0$ and the final position $X_f \gtrsim X_0 - \Delta x \approx -22a_0$ is thus always still at least $3R$ away from the bottom wall located at $x=-36a_0$. We
 measure time- and ensemble-averaged quantities for the last $50\%$ of the simulation time where the system has reached a quasi steady-state.

To simulate the dynamics of the coupled system we use a hydrid MPCD-MD scheme \cite{Gompper2009,Zottl2018}.
In the streaming step fluid particles move according to Eq.~(\ref{Eq:str}) for a time $\delta t$. Positions and velocities of the  polymer  beads evolve through molecular dynamics (MD) using a Velocity Verlet algorithm \cite{Allen1989} with time step $\delta t_P = \delta_t/10$ (except for
$k_b=3000k_BT$
where
$\delta t_P = \delta_t/50$), and the forces from the potentials [Eqs.~(\ref{Eq:Bond}) - (\ref{Eq:WCA})].
The squirmer position and orientation dynamics evolves using  a Velocity Verlet algorithm with time step  $\delta t_S = \delta t$. 
When fluid particles overlap with the squirmer or the walls a bounce-back rule is applied
where the squirmer surface velocity $\mathbf{v}_s$ has to be included and
linear and angular momenta are
exchanged accordingly \cite{Downton2009a,Zottl2018}.
This ensures the desired no-slip boundary condition and correct angular dynamics of the squirmer \cite{Goetze2010,Zottl2018}.
We consider non-adsorbing polymers, hence
squirmers and polymers interact with each other through a purely repulsive WCA potential similar as in Eq.~(\ref{Eq:WCA}) but with $\sigma^\ast=(R+\sigma/2)/2^{1/6}$ and $\epsilon_0$ is replaced by $\epsilon_S=1000k_BT$
to model
almost inpenetrable squirmers.
In the collision step not only fluid particles exchange momentum, see Eq.~(\ref{Eq:col}), but polymer  beads are coupled to the fluid by including them in the collision step \cite{Malevanets2000,Gompper2009}. Furthermore, virtual particles inside the squirmers contribute to the collision step to better  resolve the local flow fields  \cite{Downton2009a,Goetze2010,Zottl2018}. For details of the simulation procedure see SI.

\begin{figure}
   \includegraphics[width=\columnwidth]{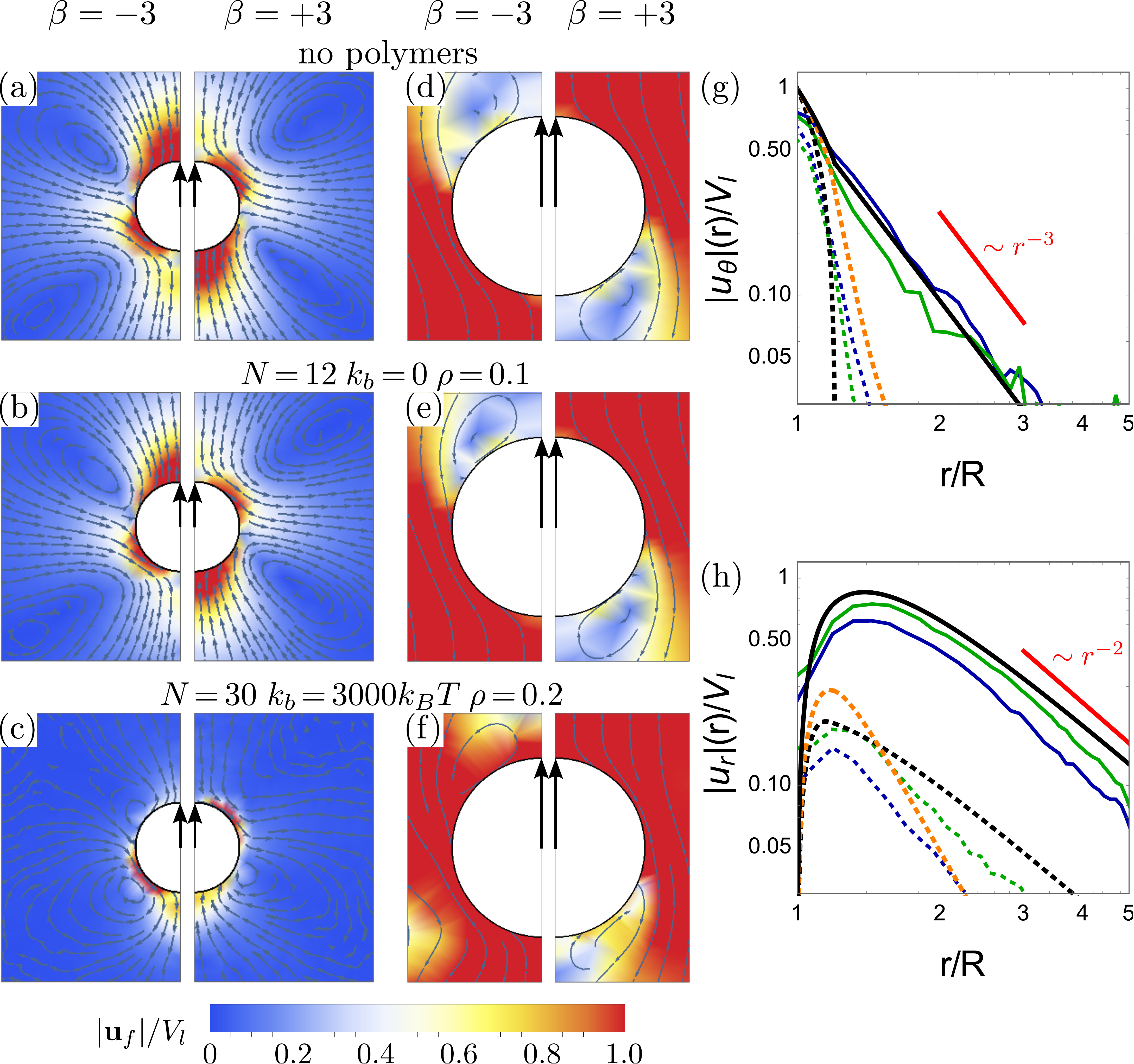}
   \caption{(a-f) Flow field around the squirmer in the lab frame (a-c) and in the co-moving frame  (d-f). (g,h) Flow-field
     decay around the squirmer equator in the tangential (g) and (h) radial direction, normalized by local tangential surface velocity $V_l$; blue: pushers, green pullers; solid lines:
     $N=12$, $k_b=0$; dotted lines: $N=30$, $k_b=3000$; black lines: from theoretical model of Ref.~\cite{Reigh2017b}; orange lines: from theoretical model of Ref.~\cite{Nganguia2020}. }
  \label{Fig:vfield}
\end{figure}

\section{Results and discussion}

\subsection{Squirmer velocity}
The steady-state squirmer velocities $V=\langle \mathbf{V}\cdot \mathbf{e} \rangle$ \cite{Downton2009a,Goetze2010,Zottl2018} are calculated from time- and ensemble averages, and are compared to the respective  velocities $V_0$ determined by simulations in the absence of polymers. We obtain $V_0=0.0406a_0/t_0$ ($\beta=0$), $V_0=0.0390a_0/t_0$ ($\beta=-3$), and $V_0=0.0409a_0/t_0$ ($\beta=+3$), which  deviate by a few percent from the theoretical value  $V_b=2B_1/3=0.0413a_0/t_0$ for a squirmer in an infinite domain at $\textnormal{Re}=0$
because of
small hydrodynamic squirmer-wall interactions  \cite{Berke2008} and small  effects from finite $\textnormal{Re}$  \cite{Wang2012b,Khair2014}.
Fig.~\ref{Fig:vf}(a) shows squirmer velocities  in solutions of short flexible polymers ($N=12$, $k_b=0$), where $V$ decreases with polymer density $\rho$ for
 all squirmer types. 
A similar trend can be observed for longer flexible polymers ($N=30$ and  $N=100$) in Fig.~\ref{Fig:vf}(b,c), but the decrease becomes stronger with increasing $N$.
Furthermore, with increasing polymer density $\rho$ pushers slow down faster compared to neutral squirmers and pullers.
The decrease of $V$ with $\rho$ and squirmer type becomes
even stronger in semiflexible polymer solutions (Fig.~\ref{Fig:vf}(d,e)).
The largest effect is observed in the stiffest filamentous solutions ($N=30$, $k_b=3000k_BT$, see Fig.~\ref{Fig:vf}(f)), where pushers become several times slower compared to pullers at high densities. Such strong deviations between pushers and pullers have so far not been predicted by continuum models.
It is thus clear that the flow microstructure and local fluid-squirmer interaction plays an important role.


Notably, we also measured the orientational squirmer dynamics and did not identify enhanced rotational diffusion of squirmers, see SI Fig S1, in contrast to recent work discussed in Ref.~\cite{Qi2020}.
  This we attribute to the relatively hard polymer-squirmer interaction and the lack of polymer adsorbtion in our work, in contrast to Ref.~\cite{Qi2020}.

\subsection{Flow field and polymer distribution}
In order to better understand our results, we measure the local fluid flow and polymer properties around the squirmer.
The flow field has been measured both in the laboratory frame of reference, and in the co-moving frame of the squirmer,
and has been averaged over ensemble, time, and along the azimuthal direction.
Examples of the  flow fields $\mathbf{u}_f(\mathbf{r})=u_\theta\hat{\boldsymbol{\theta}}+u_r\hat{\mathbf{r}_s}$ in the lab frame are shown in Fig.~\ref{Fig:vfield}(b,c),
and are compared to the flow fields in the absence of polymers [Fig.~\ref{Fig:vfield}(a)].
   Pusher and puller
   flow fields in solutions of short flexible polymers ($N=12$, $k_b=0$, $\rho=0.1$)  [Fig.~\ref{Fig:vfield}(b)] are very similar to polymer-free velocity fields.
  The decay of $|u_r|(r)$ and $|u_\theta|(r)$ along the equator of the squirmer ($\theta=\pi/2$) is shown in Fig.~\ref{Fig:vfield}(g,h) (solid blue and green curves),
and
is indeed  very similar to the polymer-free case 
decaying as $\sim r^{-2}$ and $\sim r^{-3}$, respectively, similar, as expected
in simple homogeneous viscous fluids \cite{Pedley2016,Zottl2016}.
Indeed we have previously demonstrated that solutions of short polymers can be approximated by Newtonian fluids of bulk viscosity $\eta > \eta_0$, even under non-equilibrium conditions
\cite{Zottl2019,Zottl2019b}.

\begin{figure}
  \begin{center}
  \includegraphics[width=\columnwidth]{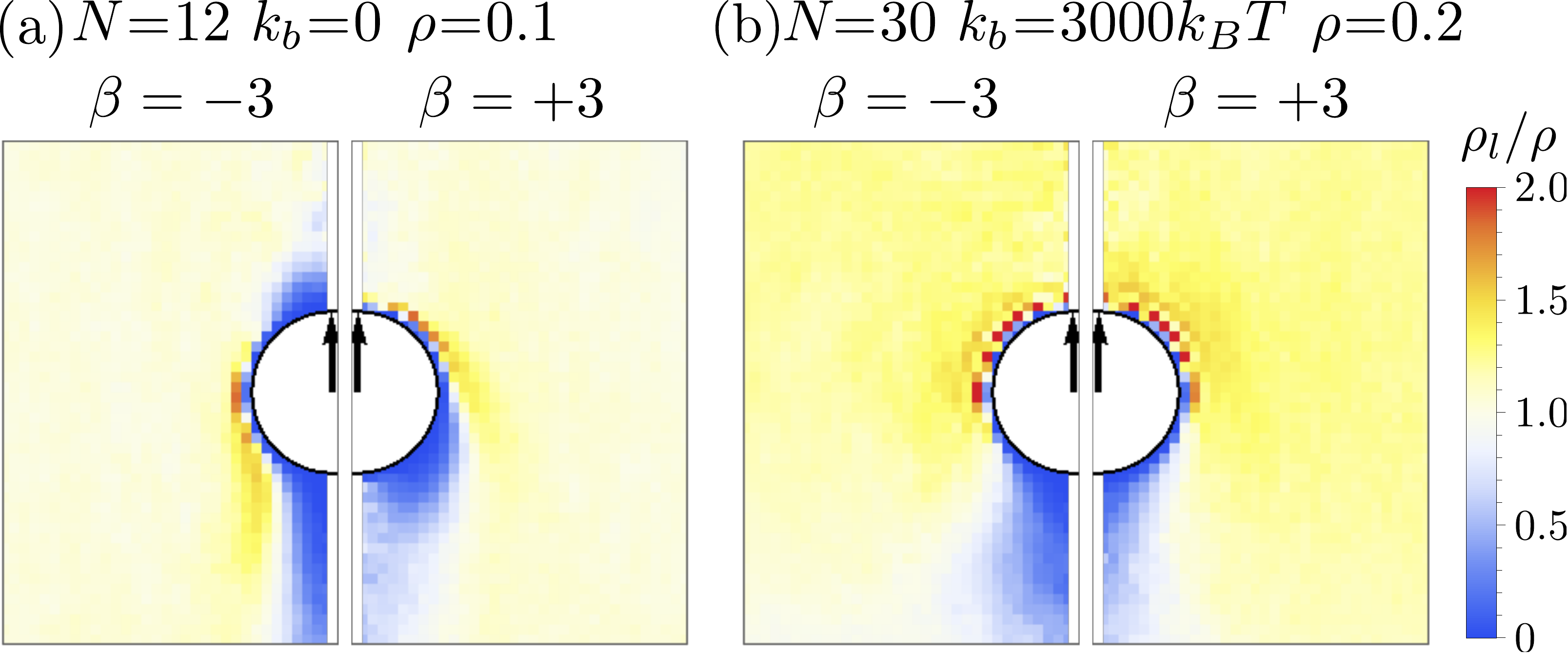}
  \caption{Local polymer density $\rho_l$ compared to bulk density $\rho$ around pushers ($\beta=-3$) and pullers ($\beta=+3$) in
    solutions of (a) short polymers and (b) long stiff filaments.
  }  
  \label{Fig:polyprop}
  \end{center}
\end{figure}

To quantify the effect of fluid viscosity we measure the $\rho$-dependent viscosity $\eta$ of all considered fluids (see SI for details), as shown in Fig.~\ref{Fig:vf}(g).
Interestingly, when we plot the squirmer velocities $V$  depending on $\eta$ [Fig.~\ref{Fig:vf}(h)] we identify a clear decrease with $\eta$, while
in continuum Newtonian fluids squirmers swim at constant velocity, independent of $\eta$ \cite{Pedley2016}.
 However, as shown in Fig.~\ref{Fig:polyprop}(a) for $N=12$, $k_b=0$, $\rho=0.1$,
 the time- and ensemble-averaged local polymer density $\rho_l$ around squirmers
is  non-homogeneous, induced by static and dynamic depletion effects
 because of finite polymer size and finite polymer diffusion time, and which depends on the squirmer type.
In a thin layer around the squirmer the polymer density is very low for all squirmer types due to polymer depletion, with a reduced viscosity around the squirmer, but the general local polymer density depends on squirmer type.

 Theoretical  two-fluid models have been used around sedimenting colloids \cite{Fan2007} and squirmers \cite{Reigh2017b},
 which are surrounded by a thin layer of thickness $\delta$ with viscosity $\eta_0$ and an ambient bulk fluid of viscosity $\eta > \eta_0$.
   Indeed it had been shown that squirmers always slow down, depending on $\eta / \eta_0$ and  $\delta / R$, independent  of the squirmer type \cite{Reigh2017b}.
   The theoretical curve from Ref.~\cite{Reigh2017b} for a single value $\delta/R = 0.2$  fits more or less all our data for neutral squirmers and pushers [Fig.~\ref{Fig:vf}(h)], in particular at sufficiently small viscosities [inset of Fig.~\ref{Fig:vf}(h)], 
   although the polymer distribution is highly non-isotropic and squirmer-type dependent  [Fig.~\ref{Fig:polyprop}(a)].
   Furthermore, the theoretical two-fluid model velocity decay fits the simulation results for sufficiently small $\eta$,
   as shown for $N=12$, $k_b=0$ in Fig.~\ref{Fig:vfield}(g,h).
   The value of $\delta/R = 0.2$ only gives an effective polymer depletion thickness, and fails to descibe pullers accurately which slow down much weaker.
   While for both pushers and pullers polymers are more depleted at the back compared to the front because of the finite diffusion time of the polymers (similar as observed around driven colloidal particles \cite{Gutsche2008,Zottl2019b}), pullers show a significant polymer accumulation in front due to the contractile flow field [Fig.~\ref{Fig:vfield}(b)].
   We speculate that this locally enhanced polymer density in front is responsible for
   pullers to be faster compared to pushers at low $\eta$, motivated by the theoretical prediction of speed enhancement for squirmers surrounded by a high-viscosity  layer for the case  $\eta_0>\eta$ \cite{Reigh2017b}.
   

As can be seen in Fig.~\ref{Fig:vf}, at larger polymer density, stiffness  (and hence larger viscosity),
  the  velocities for different squirmer types can deviate strongly.
  In these cases, also the flow fields are considerably weaker compared to the low-viscosity polymer solutions,
  as we show in Fig.~\ref{Fig:vfield}(c)  for stiff filamentous solutions ($N=30$, $k_b=3000k_BT$) at  density $\rho=0.2$, i.e.\ at viscosity $\eta=14.1\eta_0$.
The decay of the flow field components $|u_\theta|(r)$ and $|u_r|(r)$ along the equator of the squirmer, see blue and green dotted lines in Fig.~\ref{Fig:vfield}(g,h), can for both pushers and pullers fitted by  the two-fluid model \cite{Reigh2017b} (black dotted lines), again by assuming $\delta / R=0.2$.
  In addition we fit the curves by a different two-fluid model from Ref.~\cite{Nganguia2020}, which again uses a layer of viscosity $\eta_0$ close to the squirmer, but the
  outer bulk fluid is described as a Brinkman fluid as a model for porous medium with screening length $\lambda$ \cite{Leshansky2009,Nganguia2018}.
  These solutions are shown in Fig.~\ref{Fig:vfield}(g,h) as orange dotted lines, for $\delta / R=0.2$ and $\lambda=0.1 R$.
All in all, while these models capture the flow fields relatively well, neither the viscous two-fluid model \cite{Reigh2017b}, nor models of a squirmer in a Brinkman medium \cite{Leshansky2009,Nganguia2018,Nganguia2020} predict deviations in swimming speed between different squirmer types.

Interestingly, we observe large velocity deviations between pushers and pullers in stiff filamentous solutions although the polymer distribution around a pusher and a puller are similar, as shown in Fig.~\ref{Fig:polyprop}(b).
It shows an additional layer of high polymer density above the low-density layer in front of the squirmer due to frequent encounters between the moving squirmers and the relatively large filaments, which we quantify below by steric squirmer-polymer interactions.

\subsection{Direct squirmer-polymer forces}
\begin{figure}
  \begin{center}
  \includegraphics[width=\columnwidth]{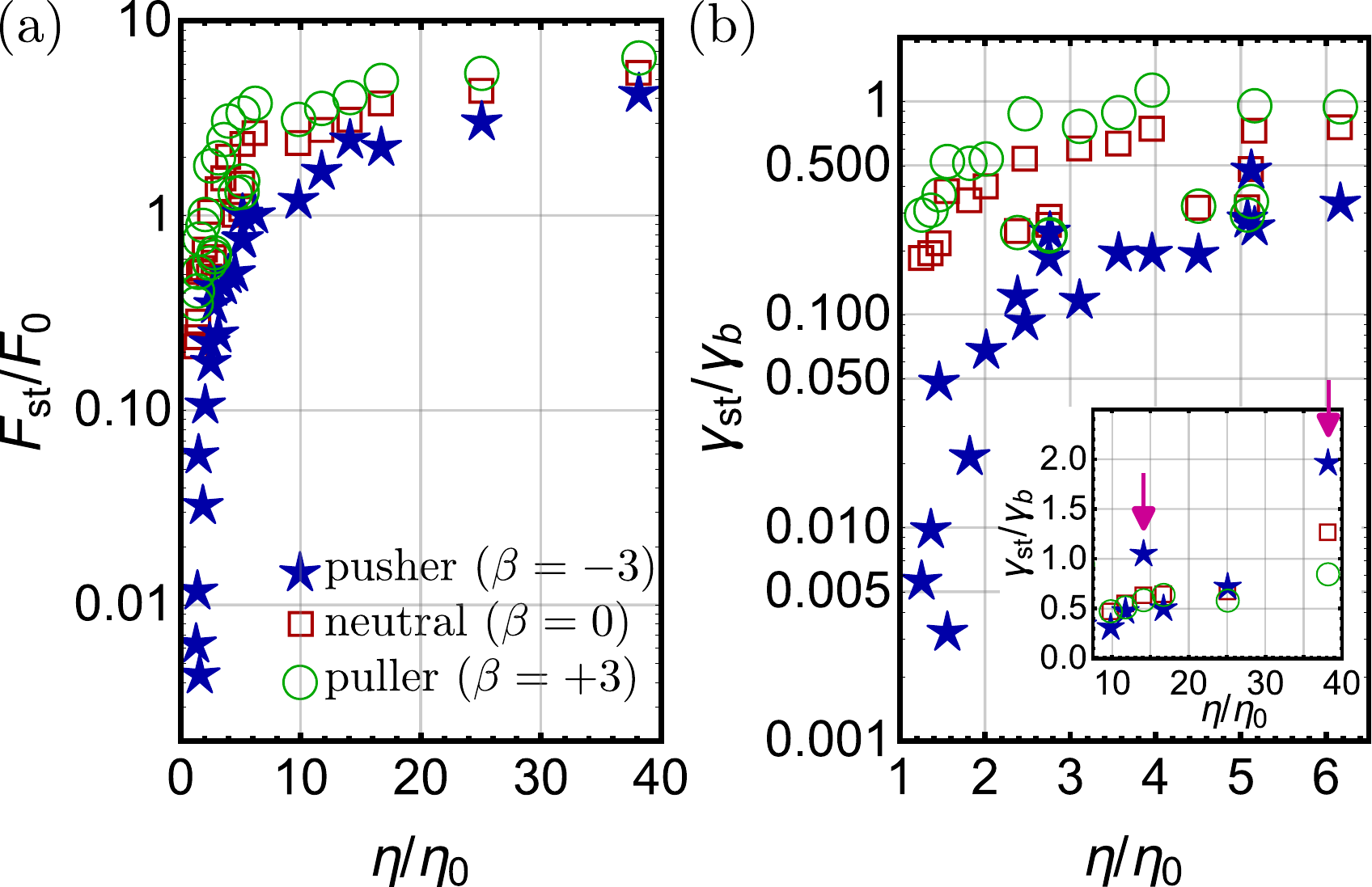}
  \caption{
    (a) Normalized forces $F_{st}$ from steric interactions between squirmers and polymers in different polymer solutions of viscosity $\eta$. (b) Effective steric friction coefficients $\gamma_{st}=F_{st}/V$ normalized by the respective Stokes bulk  friction coefficients $\gamma_b=6\pi\eta R$ for small $\eta$ (and for large $\eta$ in inset). Pink arrows indicate stiff filaments at $\rho=0.2$ and $0.3$, respectively.
  }
  \label{Fig:Fgamma}
  \end{center}
\end{figure}

Our explicit approach not only gives access to local polymer distribution around the squirmers, but also allows to measure   steric forces $\mathbf{F}_{st}=\sum_i\boldsymbol{\nabla}_iV_{WCA}(|\mathbf{r}_i-\mathbf{R}|)$ between polymers and squirmers, induced by the overlap potential  Eq.~(\ref{Eq:WCA}), which contribute in addition to  pure hydrodynamic effects to the squirmer velocities.
We plot the  time- and ensemble averaged  forces $F_{st}=-\langle \mathbf{F}_{st} \cdot \mathbf{e}\rangle$  for different fluids depending on viscosity $\eta$ in Fig.~\ref{Fig:Fgamma}(a),
normalized by the bulk Stokes force $F_0=6\pi\eta R V_b\approx 50.0k_BT/a_0$ of a sphere at velocity $V_b$ in a polymer-free fluid of viscosity $\eta_0$.
Note that $F_{st}>0$, i.e.\ steric forces act against squirmer direction $\mathbf{e}$, as expected.
For sufficiently small  viscosities $\eta$ pushers experience only a very small steric hindrance, as a consequence of the extensile flow field [Fig.~\ref{Fig:vfield}] pushing polymeric material away in front of the squirmer, while   pullers accumulate  more polymers in front because of the contractile flow field leading  to much stronger steric forces.
  See also the Supplementary Movies M1 (pusher) and M2 (puller) demonstrating the local polymer dynamics in the reference frame of the squirmer.
  Since pullers move faster compared to pushers at low viscosities [inset of Fig.~\ref{Fig:vf}(h)] we conclude that the effect on the swimming speed due to different local polymer density [Fig.~\ref{Fig:polyprop}(a)] of pullers (high density in front) compared to pushers (high density at side)  outplays the stronger steric hindrance of pullers compared to pushers at small and moderate viscosities.

To identify the effective steric friction $\gamma_{st}$ experienced by the squirmers due to direct interactions with polymers we calculate $\gamma_{st} = F_{st}/ V$ which we compare to the Stokes bulk friction coefficient in the respective fluid of viscosity $\eta$, $\gamma_b=6\pi\eta R$.
  Fig.~\ref{Fig:Fgamma}(b) shows $\gamma_{st}/ \gamma_b$ for relatively small $\eta/ \eta_0$, reiterating the fact that pushers experience higher steric friction compared to pullers.
  The inset of Fig.~\ref{Fig:Fgamma}(b) shows the steric friction for large $\eta/ \eta_0$, and we highlight results at two particular viscosities which correspond to squirmers moving in stiff filamentous solutions ($N=30$, $k_b=3000k_BT$) at respective densities $\rho=0.2$ and $\rho=0.3$
  where pushers move much slower compared to pullers [see Fig.~\ref{Fig:vf}(f)].
  This can be explained by the fact that here the steric friction for pushers is about twice as large as compared to pullers, in stark contrast as observed at lower densities and particularly when compared to (semi-)flexible polymers (see also SI Fig.~S2).
  We visualize the dynamics of stiff filaments around swimming pushers and pullers in Supplementary Movies M3 (pusher) and M4 (puller).
  While short flexible polymers (M1 and M2) easily pass by squirmers and act as large tracers in the surrounding flow,
the longer stiffer filaments are less mobile (M3 and M4).
Due to their size, individual filaments occupy larger regions around the squirmer, and do not change their conformations much because of their stiffness, see also local end-to-end distance in SI Fig.~S3.
Stiff filaments  accumulate much stronger and stay longer in front of the pusher (M3), compared to the puller (M4).
This can be explained by the flow fields in the reference frame of the squirmer, as shown
in Fig.~\ref{Fig:vfield}(d-f).
While filaments in front of pullers experience a clear sidewise flow,
pushers develop a vortex in front, as already known for polymer-free squirmers \cite{Zottl2016,Zottl2023}.
For filaments located in different parts of the vortex their center-of-mass velocity
cancels out in contrast
 to the polymer velocities in the more unidirectonal flows in front of pullers.
This leads to more frequent encounters, reorientation of filaments perpendicular to the squirmer (see also SI Fig.~S3), strong steric hindrance and hence to a strongly reduced swimming speed for pushers compared to pullers and neutral squirmers, for which flow vortices in the front  are absent.

In general, in a simplified way we can interpret the steric friction as an effective load of effective radius $R_{st}=\gamma_{st}/(6\pi \eta)$ the squirmer pushes (mainly) in front of it while swimming.
  In most of the cases this load is small, $\gamma_{st}<\gamma_{b}$ and hence $R_{st} < R$, while for example for a pusher in stiff filamentous fluids the effective load can be larger than the squirmer ($\gamma_{st}>\gamma_{b} \rightarrow  R_{st} > R$), see inset Fig.~\ref{Fig:Fgamma}(b).


Finally, we note that we also measured  stretching of flexible polymers when moving around the squirmer.
  However, this effect is small  (see also SI Fig.~S3)
  Also, while shear-thinning effects are known to create local viscosity gradients and slowing down of squirmers, this effect is  independent of squirmer type \cite{Datt2015xx}.
  Furthermore, since our polymers only consist of a relatively small number of monomers ($N \le 100$), entanglement effects are expected to be negligible \cite{Zottl2019}.
  Notably, even when viscoelastic effects are accounted for, the effect on squirmer speed is typically small \cite{Zhu2012}.

\section{Conclusion}
Our work demonstrates the importance of non-continuum effects on the dynamics of different squirmers in fluids consisting of large polymers and filaments.
  First, non-homogeneous local polymer densities lead to non-homogeneous and squirmer-type specific viscosity gradients, which slow down squirmers but can only partly be captured by effective two-fluid models.
    How the specific polymer density landscape affects the squirmer speed remains to be investigated in more detail by refined theoretical models in future work.
  Second, because of their large size, polymers at sufficiently large size and stiffness act as an effective load where the squirmer pushes against, quantified by an effective steric friction.
In this sense our approach bridges the gap between available continuum hydrodynamic models and purely non-hydrodynamic models of active Brownian particles moving in polymer solutions \cite{Du2019,Kim2022}, by
capturing non-continuum effects of squirmer hydrodynamics in complex fluids.

Finally we want to note that our results are expected to be relevant for real biological  microswimmers moving in solutions of polymers and filaments which are  comparable in size with the microswimmer itself. Examples are bacteria  or sperm cells  moving in mucus \cite{Li2021b,Figueroa-Morales2019}, where both the hydrodynamic and steric effects are expected to be of relevance for their transport in these heterogeneous fluid environments.


\acknowledgments
AZ acknowledges funding from the Austrian Science Fund (FWF) through a Lise-Meitner Fellowship (Grant No M 2458-N36).
The computational results presented have been achieved  using the Vienna Scientific Cluster (VSC).


\newpage

\newcommand{\az}[1]{{\color{blue}#1\color{black}}}
\newcommand{\azz}[1]{{\color{black}#1\color{black}}}

\newcommand{\eh}{\mathbf{e}}
\newcommand{\deriv}[2]{\frac{\mathrm{d} #1}{\mathrm{d} #2}}
\newcommand{\hvec}[1]{\hat{\mathbf{#1}}}

\newcommand{\tEins}{\textbf{\textsf{1}}}

\newcommand{\Pe}{\mathrm{Pe}}
\newcommand{\Per}{\mathrm{Pe}_r}

\newcommand{\vhh}[1]{\hat{\mathbf{#1}}}
\newcommand{\hsy}[1]{\hat{\boldsymbol{#1}}}
\newcommand{\RCh}{R_{\mathrm{Ch}}}
\newcommand{\RSq}{R_{\mathrm{Sq}}}
\newcommand{\vdd}[1]{\frac{d}{dt}\mathbf{#1}}
\newcommand{\vdh}[1]{\frac{d}{dt}\hat{\mathbf{#1}}}

\newcommand{\hphi}{\hat{\boldsymbol{\varphi}}}
\newcommand{\hrho}{\hat{\boldsymbol{\rho}}}
\newcommand{\hz}{\hat{\mathbf{z}}}
\newcommand{\ephi}{e_\varphi}
\newcommand{\erho}{e_\rho}
\newcommand{\ez}{e_z}
\newcommand{\htheta}{\hat{\boldsymbol{\theta}}}
\newcommand{\vf}{\bar{v}_f}
\newcommand{\HDD}{H_\textnormal{2D}}
\newcommand{\HDDD}{H_\textnormal{3D}}

\newcommand{\refEq}[1]{Eq.~(\ref{Eq:#1})}
\newcommand{\refEqu}[1]{Eqs.~(\ref{Eq:#1})}
\newcommand{\refFig}[1]{Fig.~\ref{Fig:#1}}
\newcommand{\refFigu}[1]{Figs.~\ref{Fig:#1}}
\newcommand{\refFigure}[1]{Figure~\ref{Fig:#1}}
\newcommand{\refFigures}[1]{Figures~\ref{Fig:#1}}
\newcommand{\refSec}[1]{Sec.~\ref{Sec:#1}}
\newcommand{\refChapt}[1]{Chapt.~\ref{Chapt:#1}}
\newcommand{\vecin}[2]{#1 \cdot #2}
\newcommand{\tensA}[1]{\boldsymbol{#1}}

\renewcommand{\ge}{\geqslant}
\renewcommand{\le}{\leqslant}

\newcommand{\nablabf}{\boldsymbol{\nabla}}

\newcommand{\ten}[1]{\vec{\sf #1}}
\newcommand{\NMD}{N_{\textnormal{MD}}}

\newcommand{\hx}{\hat{\vec{x}}}
\newcommand{\hy}{\hat{\vec{y}}}
\newcommand{\const}{\text{const}}

\newcommand{\vout}{\vec{v}_\text{out}}
\newcommand{\vin}{\vec{v}_\text{in}}
\newcommand{\viout}{\vec{v}_\text{i,out}}
\newcommand{\viin}{\vec{v}_\text{i,in}}

\setcounter{figure}{0} 
\renewcommand{\thefigure}{S\arabic{figure}}

\onecolumn

\noindent {\Large Supplementary Information: \\ Dynamics of  squirmers in explicitly modeled  polymeric fluids}\\[4mm]


Here we present the details of the MPCD simulations of a squirmer moving in a solution of polymers. The detailed description of the motion of fluid particles and squirmers is based on Refs.~\cite{Zottl2014b,Zottl2018}.

\section{Initial setup}
The system  consists of (i) a fixed number of $N_f$ pointlike MPCD fluid particles of mass $m$ at positions $\mathbf{x}_i$ and velocities $\mathbf{v}_i$, $i=1,\dots,N_f$, of (ii) $N_p$ polymers consisting each of $N$
monomers
of diameter $\sigma$, mass $m_p$, positions  $\mathbf{r}_i$ and velocities $\mathbf{w}_i$, $i=1,\dots,N N_p$, of (iii) a squirmer of radius $R$ and surface velocity modes $B_1$ and $\beta$ with position $\mathbf{R}$, orientation $\mathbf{e}$, velocity $\mathbf{V}$
and angular velocity $\boldsymbol{\Omega}$, and of (iv) two solid walls located at $\pm S_X/2$.
We employ periodic boundary conditions in $y$ and $z$ direction.
The toal simulation volume $\mathcal{V}=S_XS_YS_Z$ is given by $S_X=72a_0$, $S_Y=48a_0$ and $S_Z=48a_0$.
The length $a_0$ is the length of the MPCD simulation box, and the basic length scale in our system.
Furthermore we employ $m$ as the basic mass scale in the system, and $k_BT$ of the energy scale of the system, where $T$ is the temperature of fluid kept constant by using a thermostat (see below).
Then the unit of time is $t_0=a_0\sqrt{m/k_BT}$.

Initially the squirmer is placed at position $\mathbf{R}_0=\{18a_0,0,0\}$, orientation $\mathbf{e}_0=\{-1,0,0\}$, velocity $\mathbf{V_0}=0$ and angular velocity $\boldsymbol{\Omega}_0=0$.
The total number of fluid particles is chosen such to fulfill an average number $n$ of particles per unit volume $a_0^3$,
and is given by $N_f=n (\mathcal{V}-\mathcal{V}_{Sq})/a_0^3$ with $\mathcal{V}_{Sq}=4\pi R^3/3$.
Two parameters define the transport properties such as the viscosity of the fluid, namely the time step $\delta t$ between two fluid collisions (see below),
and the value of $n$.

The squirmer is assumed to be neutrally buoyant, i.e. it has the same mass density as the fluid, $\rho_{Sq}=\rho_f=nma_0^{-3}$,
and its mass $M_S$ is given by $M_S=\rho_{Sq}\mathcal{V}_{Sq}$, and its moment of inertia by $I_S=\frac{2M_S R^2}{5} = \frac{8\pi n R^5}{15} \frac{m}{a_0^3}$.


Initially, the effective fluid particles are randomly dis\-tri\-buted in the simulation domain but are not allowed to overlap with the squirmer.
The fluid velocities $\mathbf{v}_i$ are Gaussian distributed with zero mean and standard deviation $\sigma_0 = \sqrt{k_BT/m}$.

The number of polymers $N_p$ in the system is given by fixing the total volume fraction $\rho=N N_p \pi \sigma^3/(6(\mathcal{V}-\mathcal{V}_{Sq}))$ of monomers in the simulation domain.
The mass of the monomers is set to $m_p=10m$ to achieve good coupling between fluid and polymers \cite{Gompper2009}.
Polymers are initially randomly distributed in the simulation domain with velocities $\mathbf{w}_i$ drawn from a Gaussian distribution with zero mean and standard deviation $\sigma_p = \sqrt{k_BT/m_p}$.

\section{Streaming step}
In the streaming step  squirmers and fluid particles  move ballistically for a time $\delta t$,
while the dynamics of the polymers is captured by a molecular dynamics (MD) scheme, as described below.

\subsection{Motion of the squirmers}
Before we integrate the equations of motion for the squirmers we have to calculate the forces acting on the squirmer through the overlap with polymer beads.
These forces are exactly the steric forces described in the main text,
$\mathbf{F}_{st}=\sum_i\boldsymbol{\nabla}_iV_{WCA}(|\mathbf{r}_i-\mathbf{R}|)$,
obtained from Eq.~(5) in the main text  but with $\sigma^\ast=(R+\sigma/2)/2^{1/6}$ and $\epsilon_0$ is replaced by $\epsilon_S=1000k_BT$.
Furthermore, the squirmer assumes a torque from the interaction with the polymers,
$\mathbf{T}_{st} = \sum_i(\mathbf{R}-\mathbf{r}_i)\times \boldsymbol{\nabla}_iV_{WCA}(|\mathbf{r}_i-\mathbf{R}|)$

We use a simple \textit{Velocity-Verlet} integration  for a time $\delta t$. It updates positions and orientations according to
 \begin{equation}
 \begin{split}
   \mathbf{R}(t+\delta t) &= \mathbf{R}(t)+ \left[ \mathbf{V}(t) + \frac{1}{2M_S} \mathbf{F}_{st}(t)\delta t  \right] \delta t, \\
   \eh(t+\delta t) &= \eh(t)+\left[ \left( \boldsymbol{\Omega}(t) + \frac{1}{2I_S} \mathbf{T}_{st}(t)\delta t \right)\times\eh(t)\right] \delta t,
 \end{split}
\label{Eq:dr2}
\end{equation}
and translational and angular velocities 
according to
 \begin{equation}
 \begin{split}
   \mathbf{V}(t+\delta t) &= \mathbf{V}(t)+ \frac{1}{2M_S}\left[ \mathbf{F}_{st}(t)+ \mathbf{F}_{st}(t+\delta t)\right] \delta t, \\
   \boldsymbol{\Omega}(t+\delta t) &= \boldsymbol{\Omega}(t)+ \frac{1}{2I_S}\left[ \mathbf{T}_{st}(t)+ \mathbf{T}_{st}(t+\delta t)\right] \delta t.
 \end{split}
\label{Eq:dr3}
\end{equation}

\subsection{Motion of the fluid particles}
In the  streaming step the fluid particles  simply move ballistically  by
\begin{equation}
  \mathbf{x}_i (t + \delta t) = \mathbf{x}_i (t) + \mathbf{v}_i(t) \delta t \, .
  \label{Eq:str}
\end{equation}
When a fluid particle hits a wall, it is moved  half a time step back and its velocity is updated
according to the bounce-back rule to fulfill the no-slip boundary condition at the wall,
where the velocity is simply reversed,
\begin{equation}
\mathbf{v}_i' = - \mathbf{v}_i \, ,
\label{Eq:fluidIA1}
\end{equation}
Then, the fluid particle is moved forward for half a time step with the new velocity.
A similar procedure is applied if a fluid particle hits the squirmer, but we 
have to account for the fact that the surface of the squirmer at position $\mathbf{r}_S$ 
contains a local velocity $\mathbf{v}_s(\eh,\mathbf{r}_s)$  and translation and rotation of the squirmer \cite{Downton2009a,Goetze2010},
\begin{equation}
  \mathbf{v}_i' = -\mathbf{v}_i + 2\left[ \mathbf{v}_s(\eh,\mathbf{r}_{i}^\ast)
  + \boldsymbol{\Omega} \times (\mathbf{r}_{i}^\ast-\mathbf{R}) + \mathbf{V} \right] \, .
\label{Eq:fluidIA2}
\end{equation}
Here,
$\mathbf{r}_{i}^\ast$ is the collision point of the particle on the surface of the squirmer
and $ \mathbf{v}_s(\eh,\mathbf{r}_{i}^\ast)$ is the 
surface velocity of the squirmer at position  $\mathbf{r}_{i}^\ast$.

When a fluid particle interacts with  a squirmer or a wall, its  momentum $\mathbf{p}_i=m\mathbf{v}_i$ is modified.
While a fixed wall absorbs this momentum, moving objects such as colloids or squirmers 
update their velocity and angular velocity such that the total momentum and angular momentum is conserved during the collision.
The change of momentum for fluid particle $i$ hitting the  squirmer is simply 
\begin{equation}
  \Delta \mathbf {p}_{i}=m[ \mathbf{v}_i' - \mathbf{v}_i ] \, ,  
\end{equation}
which is then transferred to the squirmer.
In general,
if $N^\ast$ fluid particles hit the squirmer
during time interval $\delta t$,
its velocity $\mathbf{V}$  and angular velocity $\boldsymbol{\Omega}$  is updated  to
 \begin{equation}
 \begin{split}
   \mathbf{V}' &= \mathbf{V} +  \frac{1}{M_S}  \Delta \mathbf{P}^s, \\
   \boldsymbol{\Omega}' &= \boldsymbol{\Omega} +\frac{1}{I_S}   \Delta \mathbf{L}^s,
 \end{split}
\end{equation}
where
 \begin{equation}
 \begin{split}
  \Delta \mathbf{P}^s &= -\sum_{i=1}^{N^\ast}\Delta \mathbf{p}_{i}, \\
  \Delta \mathbf{L}^s &= - \sum_{i=1}^{N^\ast} \left[ ( \mathbf{r}_{i}^\ast - \mathbf{R}) \times \Delta\mathbf{p}_{i} \right].
 \end{split}
\label{Eq:dPStr}
\end{equation}
are
the total  linear and angular momentum 
transferred to the  squirmer in the streaming step by the fluid particles.

\subsection{Motion of the polymers}
The polymer beads move by molecular dynamics with time step $\delta t_P = \delta_t/10$ (except for
$k_b=3000k_BT$ where $\delta t_P = \delta_t/50$),
and the monomers interact with each other and with the squirmer through the potentials Eq.~(3)-(5) in the main text.
A total force $\mathbf{F}_i$ on each monomer is calculated in each time step as the gradient of the potentials.
Then their positions and velocities are updated according to  
 \begin{equation}
   \mathbf{r}_i(t+\delta t_p) = \mathbf{r}_i(t)+ \left[ \mathbf{w}_i(t) + \frac{1}{2m} \mathbf{F}_i(t)\delta t_p  \right] \delta t_p
\end{equation}
and
 \begin{equation}
 \begin{split}
   \mathbf{w}_i(t+\delta t_p) = \mathbf{w}_i(t)+ \frac{1}{2m}\left[ \mathbf{F}_{i}(t)+ \mathbf{F}_{i}(t+\delta t_p)\right] \delta t_p \, .
 \end{split}
\label{Eq:dr3}
\end{equation}

\section{Collision step}
In the collision step fluid particles interact with each other but also interact with the squirmer and the walls 
via \textit{virtual particles} \cite{Lamura2001,Gompper2009},
and with polymer beads.
In order to fulfill Galilean invariance but also to reduce memory effects and correlations in the collision step,
the cell grid is first shifted randomly at each time step  by a random vector $\mathbf{s}=\{s_1,s_2,s_3\}$, where the components $s_i$ are drawn from a uniform distribution $s_i \in [-a_0/2,a_0/2]$.
Then, collision cells partly overlapping with squirmers and walls are filled with virtual particles, which are placed in the walls and in the squirmer.
This increases the accuracy of the hydrodynamic flow fields significantly since otherwise these cells would have an average fluid particle number below the mean number $n$
and hence  locally a smaller viscosity \cite{Zottl2018}.

The virtual particles are placed  randomly distributed  at the same density as the fluid, $\rho_f$, in a layer of thickness $a_0$ inside a bounding wall and in layers of thickness $\sqrt{3}a_0$ inside the squirmers.
It is thus guaranteed that grid cells overlapping with walls and squirmers are always completely filled.
The positions and velocities
of the  virtual particles are denoted 
by  $\bar{\mathbf{x}}_{i}$  and $\bar{\mathbf{v}}_{i}$, 
respectively,
with $i=1,\dots, N^v$ and $N^v$ is the total number of virtual particles. 
Their velocities are drawn 
 from a normal distribution with 
 the usual
standard deviation, $\sigma_0=\sqrt{k_BT/m}$, and zero mean.
In addition, virtual particles 
located inside the squirmer  also assume the local 
surface velocity of the point $\bar{\mathbf{r}}_{i}^\ast$  on the squirmer surface, which is closest to $\bar{\mathbf{x}}_i$,
plus the velocity of this point due to the translation and rotation of the squirmer
\cite{Goetze2010}:
\begin{equation}
\bar{\mathbf{v}}_{i} = \bar{\mathbf{v}}_{i}^\text{r} + \mathbf{v}_s(\eh,\bar{\mathbf{r}}_{i}^\ast)
  + \boldsymbol{\Omega} \times (\bar{\mathbf{r}}_{i}^\ast-\mathbf{R}) + \mathbf{V},
\end{equation}
where $\bar{\mathbf{v}}_{i}^\text{r}$ are the random velocities.

Then fluid particles, virtual particles and the monomers are sorted into the cells, which all take part in the collision step.
We
denote the number of fluid, virtual and monomer particles in cell $\xi$ by $N_\xi^f$, $N_\xi^v$ and $N_\xi^m$, respectively.
The total number of particles in the cell is $N_\xi=N_\xi^f+N_\xi^v+N_\xi^m$
where the total number of monomers in a cell, $N_\xi^m$,  is typically either zero or one since $\sigma=a_0$.
In order to perform the collision step, the mean velocity in a cell is computed,
\begin{equation}
 \mathbf{u}_\xi = \frac{1}{mN_\xi^f+mN_\xi^v+m_pN_\xi^m}\left(m\sum_{i=1}^{N_\xi^f}\mathbf{v}_i + m\sum_{i=1}^{N_\xi^v}\bar{\mathbf{v}}_i + m_p\sum_{i=1}^{N_\xi^m}\mathbf{w}_i \right) \, .
\end{equation}
We use an Anderson thermostat \cite{Noguchi2007,Noguchi2008},
where random  velocities for all fluid particles, virtual particles and monomers have to be computed, which are denoted by $\mathbf{v}_i^{\text{r}}$, $\bar{\mathbf{v}}_i^{\text{r}}$ and $\mathbf{w}_i^{\text{r}}$, respectively, 
and which are
again drawn from a Gaussian distribution with
width $\sigma_0=\sqrt{k_BT/m}$ for fluid and virtual particles and $\sigma_p=\sqrt{k_BT/m_p}$ for the monomers.
To conserve linear momentum the change of total velocity due to the added random velocities, 
$\boldsymbol{\mathcal{V}}_\xi$, has to be calculated,
given by
\begin{equation}
\boldsymbol{\mathcal{V}}_\xi = \frac{1}{mN_\xi^f+mN_\xi^v+m_pN_\xi^m}\left(m\sum_{i=1}^{N_\xi^f}\mathbf{v}_i^r + m\sum_{i=1}^{N_\xi^v}\bar{\mathbf{v}}_i^r + m_p\sum_{i=1}^{N_\xi^m}\mathbf{w}_i^r \right) \, .
\end{equation}
Next,
the center of mass has to be calculated,
\begin{equation}
 \mathbf{r}_\xi^s = \frac{1}{mN_\xi^f+mN_\xi^v+m_pN_\xi^m}\left(m\sum_{i=1}^{N_\xi^f}\mathbf{x}_i + m\sum_{i=1}^{N_\xi^v}\bar{\mathbf{x}}_i + m_p\sum_{i=1}^{N_\xi^m}\mathbf{r}_i \right) \, ,
\end{equation}
and the relative positions $\mathbf{x}_i^s = \mathbf{x}_i - \mathbf{r}_\xi^s$,
$\bar{\mathbf{x}}_i^s = \bar{\mathbf{x}}_i - \mathbf{r}_\xi^s$ and $\mathbf{r}_i^s = \mathbf{r}_i - \mathbf{r}_\xi^s$.
They are needed to calculate
the inverse of the moment of inertia tensor $\mathbf{I}^{-1}_\xi$ in each cell,
where
\begin{equation}
  \mathbf{I}_\xi = m  \sum_{i=1}^{N_\xi^f}(|\mathbf{x}_i^s|^2\,\tEins - \mathbf{x}_i^s\otimes\mathbf{x}_i^s)
         + m\sum_{i=1}^{N_\xi^v}(|\bar{\mathbf{x}}_i^s|^2\,\tEins - \bar{\mathbf{x}}_i^s\otimes\bar{\mathbf{x}}_i^s) + m_p  \sum_{i=1}^{N_\xi^m}(|\mathbf{r}_i^s|^2\,\tEins - \mathbf{r}_i^s\otimes\mathbf{r}_i^s).
\end{equation}
The random velocities added in the collision step change angular momentum by
\begin{equation}
  \Delta \boldsymbol{\mathcal{L}}_\xi = m \sum_{i=1}^{N_\xi^f}[\mathbf{x}_i^s\times(\mathbf{v}_i-\mathbf{v}_i^\text{r})]   
   +  m \sum_{i=1}^{N_\xi^v}[\bar{\mathbf{x}}_i^s\times(\bar{\mathbf{v}}_i-\bar{\mathbf{v}}_i^\text{r})] + m_p \sum_{i=1}^{N_\xi^m}[\mathbf{r}_i^s\times(\mathbf{w}_i-\mathbf{w}_i^\text{r})]     .
\end{equation}
To compensate for $ \Delta \boldsymbol{\mathcal{L}}_\xi$ and
conserve angular momentum in each cell, 
the following angular velocity 
is computed \cite{Noguchi2007},
\begin{equation}
 \boldsymbol{\omega}_\xi^{\text{AMC}} = \mathbf{I}_\xi^{-1}  \Delta \boldsymbol{\mathcal{L}}_\xi,   
\label{Eq:AT+aOm}
\end{equation}
and used
to rotate 
the
particle velocities in a cell.
Thus, by adding the
extra terms $\boldsymbol{\omega}_\xi^{\text{AMC}} \times \mathbf{x}_i^s$,
 $ \boldsymbol{\omega}_\xi^{\text{AMC}} \times \bar{\mathbf{x}}_i^s$ or $\boldsymbol{\omega}_\xi^{\text{AMC}} \times \mathbf{r}_i^s$
to the new  velocities, angular momentum is conserved without changing linear momentum.
To summarize, in the collision step the particle
velocities are updated according to \cite{Noguchi2007}
\begin{equation}
 \begin{split}
  \mathbf{v}_i' &=  \mathbf{u}_\xi +  \mathbf{v}_i^{\text{r}} - \boldsymbol{\mathcal{V}}_\xi + \boldsymbol{\omega}_\xi^{\text{AMC}} \times \mathbf{x}_i^s, \\
  \bar{\mathbf{v}}_i' &= \mathbf{u}_\xi +  \bar{\mathbf{v}}_i^{\text{r}} - \boldsymbol{\mathcal{V}}_\xi  + \boldsymbol{\omega}_\xi^{\text{AMC}} \times \bar{\mathbf{x}}_i^s \, , \\
  \mathbf{w}_i' &=  \mathbf{u}_\xi +  \mathbf{w}_i^{\text{r}} - \boldsymbol{\mathcal{V}}_\xi + \boldsymbol{\omega}_\xi^{\text{AMC}} \times \mathbf{r}_i^s. \\
 \end{split}
\label{Eq:vAMC+a}
\end{equation}

Then momentum and angular momentum is transferred to the squirmers:
The change of  momentum for a virtual particle is 
 \begin{equation}
  \Delta \bar{\mathbf{p}}_i= 
    m(\bar{\mathbf{v}}_i'-\bar{\mathbf{v}}_i) .  
\end{equation}
The momenta and angular momenta of all
virtual particles  are then added
up and assigned
to the squirmer.
Thus, after a collision step the
squirmer assumes an additional momentum 
$ \Delta \mathbf{P}^c$ and angular momentum $ \Delta \mathbf{L}^c$,
 \begin{equation}
 \begin{split}
  \Delta \mathbf{P}^c &= \sum_{i=1}^{N_S^v}\Delta \bar{\mathbf{p}}_i, \\
  \Delta \mathbf{L}^c &= \sum_{i=1}^{N_S^v} ( \bar{\mathbf{r}}_i - \mathbf{R}) \times \Delta\bar{\mathbf{p}}_i,  \\
 \end{split}
\label{Eq:dPCol}
\end{equation}
where the sum goes over all virtual particles   located in the squirmer with total number $N_S^v$.
So, 
after each collision step
the squirmer velocity and angular velocity are updated according to
 \begin{equation}
 \begin{split}
   \mathbf{V}' &= \mathbf{V} +\frac{1}{M_S} \Delta \mathbf{P}^c, \\
   \boldsymbol{\Omega}' &= \boldsymbol{\Omega} +\frac{1}{I_S} \Delta \mathbf{L}^c \, .
 \end{split}
\end{equation}

\section{Viscosity Measurements with MPCD Poiseuille flow}
We measure the viscosity $\eta$ of all fluid similar as in our previous work \cite{Zottl2019}.
It follows the method proposed in Ref.~\cite{Backer2005} for Newtonian fluids in the absence of polymers.
We use a system size of $S_X=60a_0$, $S_Y=S_Z=30a_0$ and use periodic boundary conditions in all three dimensions and fluid particles and polymers are initially placed randomly.

Then the fluid  particles are subjected to a constant but small acceleration force $f_a=10^{-3}k_BT/a_0$  \cite{Allahyarov2002}  in $z$ direction in one half of the simulation box ($x>0$) and to $f_a=-10^{-3}k_BT/a_0$ in $z$ direction in the other half ($x<0$).
Then the streaming step has to be modified and positions and velocities of the fluid particles are updated according to,
\begin{equation}
  \begin{split}
\mathbf{x}_i(t+\delta t) &= \mathbf{x}(t) + \mathbf{v}_i(t)\delta t + \frac{1}{2m} \mathbf{f}_a \delta t^2, \\
\mathbf{v}_i(t+\delta t) &= \mathbf{v}_i(t) + \frac{1}{m}  \mathbf{f}_a \delta t.
\end{split}
\label{Eq:StrF}
\end{equation}
where $\mathbf{f}_a=f_a\hat{\mathbf{z}}$.
Even in the absence of any walls this results in a periodic Poiseuille flow profile
for sufficiently small $f_a$, i.e.\ shear-thinning is still negligible.
The fitted maximum velocity $v_{max}$ of the quadratic flow profile, averaged over time and 16 simulation runs, is linearly related to the inverse viscosity $\eta^{-1}$ \cite{Backer2005,Zottl2019},
\begin{equation}
\eta = \frac{n f_a S_X^2}{8a_0^3v_{max}}
\end{equation}
In the absence of polymers we measure $v_{max}^0=0.070a_0/t_0$ which corresponds to viscosity $\eta_0=16.04\sqrt{m k_BT /a_0^4}$, as obtained in previous work \cite{Zottl2012,Zottl2014b}.
In the presence of polymers $v_{max}<v_{max}^0$ in the presence of polymers which enables us to determine all the viscosities $\eta > \eta_0$ presented in Fig.~2(g) in the main text.

\section{Supplementary Figures}
\subsection{Rotational diffusion}
\begin{figure}[h]
   \includegraphics[width=\columnwidth]{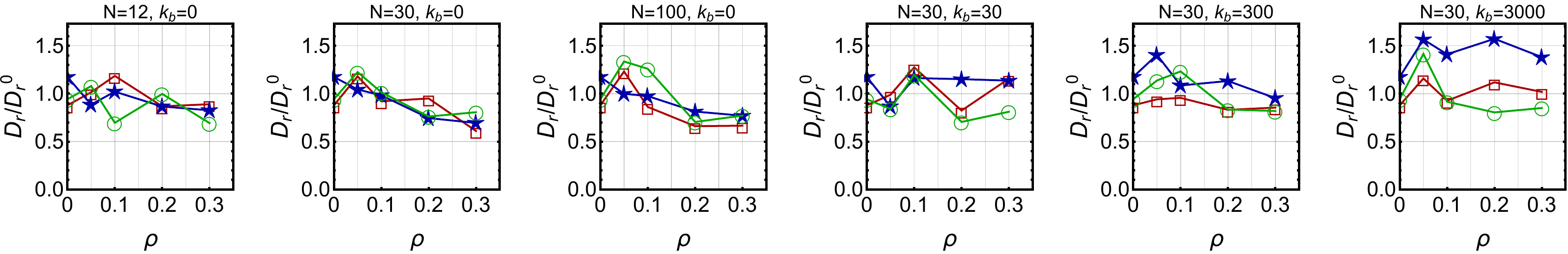}
   \caption{Measured rotational diffusion constant $D_r$ compared to the theoretical value in the absence of polymers $D_r^0$, for different fluids at different densities $\rho$. blue: pushers, green: pullers, red: neutral squirmers.}
  \label{Fig:S1}
\end{figure}

We  measure the orientational correlation $C_e(t)$ of the squirmer,
$C_e(t)= \langle \mathbf{e}(0)\cdot \mathbf{e}(t) \rangle$, where time $t=0$ is set after $50\%$ of the simulation time.
From $C_e(t)$ we fit the theoretical curve  $C_e^{th}(t)=e^{-2D_r t}$ to obtain the
rotational diffusion constants $D_r$.
We show the results in Fig.~\ref{Fig:S1} where we compare $D_r$ to the theoretical value in the absence of polymers, $D_r^0=k_BT/(8\pi\eta_0 R^3)$.

\subsection{Steric friction}
\begin{figure}[h]
   \includegraphics[width=\columnwidth]{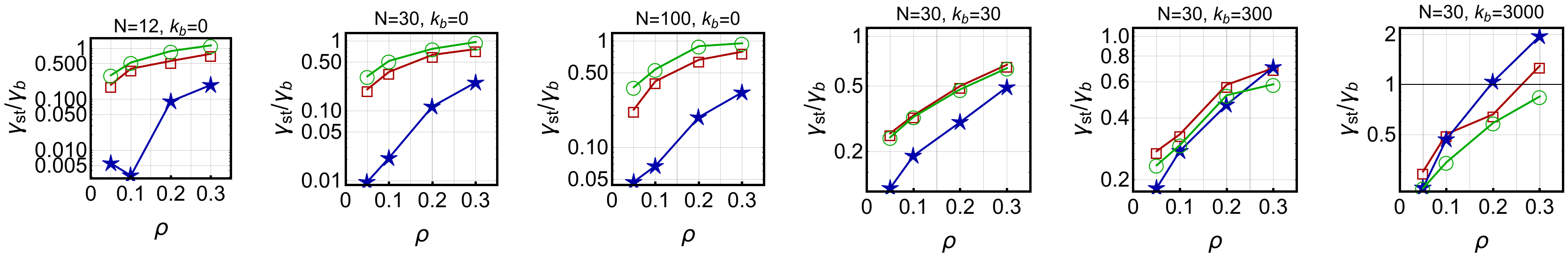}
   \caption{Effective steric friction coefficient $\gamma_{st}$  for different polymer types depending on density. The data points are the same as used in Fig.~5(b) in the main text.}
  \label{Fig:S2}
\end{figure}

We plot the effective steric friction coefficient $\gamma_{st}$ devided by the bulk friction coefficient $\gamma_b$, similar as in Fig.~5(b) in the main text but now depending on polymer length, stiffness and density in Fig.~\ref{Fig:S2}.

\subsection{Local polymer properties}
\begin{figure}[h]
   \includegraphics[width=\columnwidth]{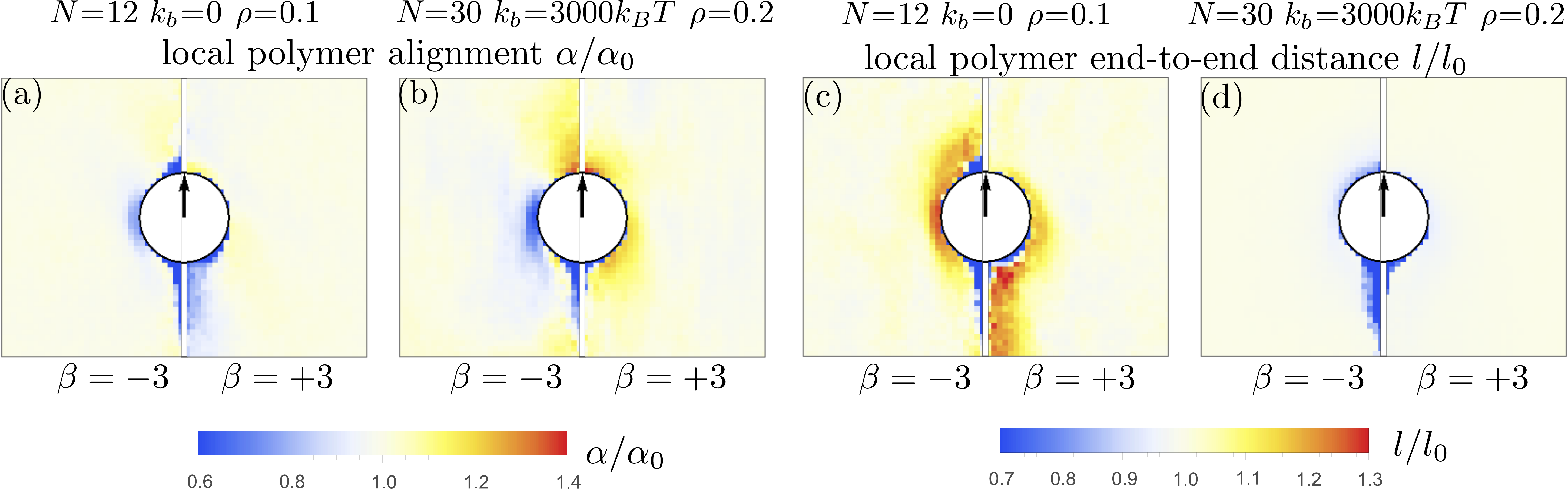}
   \caption{(a,b) Local polymer orientation angle $\alpha$ compared to the average bulk value $\alpha_0$. (c,d) Local polymer end-to-end distance compared to the average value $l_0$. Shown are results for differents quirmer types and different fluids.}
  \label{Fig:S3}
\end{figure}

In Fig.~\ref{Fig:S3}(a,b) we show the local polymer alignment angle $\alpha$ compared to the
average angle $\alpha_0$ defined as the angle between the polymer end-to-end vector and the squirmer orientation for two different fluids.
Here  $\alpha=0$ corresponds to alignement in the direction of the squirmer, and $\alpha=\pi/2$ perpendicular to it.

In Fig.~\ref{Fig:S3}(c,d) we show the local polymer end-to-end distance $l$ compared to the average value $l_0$ for the same two fluids, which indicates regions of small stretching (typically $l/l_0<1.3$) for flexible polymers and small compression for stiff filaments.

\section{Supplementary Movies}
\subsection{Movie M1}
M1 shows the dynamics of a pusher ($\beta=-3$) and its local polymeric environment, characterized by polymer length $N=12$, stiffness $k_b=0$ and density $\rho=0.1$, in the reference frame of the moving squirmer.
\subsection{Movie M2}
M2 shows the dynamics of a puller ($\beta=+3$) and its local polymeric environment, characterized by polymer length $N=12$, stiffness $k_b=0$ and density $\rho=0.1$, in the reference frame of the moving squirmer.
\subsection{Movie M3}
M3 shows the dynamics of a pusher ($\beta=-3$) and its local polymeric environment, characterized by polymer length $N=30$, stiffness $k_b=3000k_BT$ and density $\rho=0.2$, in the reference frame of the moving squirmer.
\subsection{Movie M4}
M4 shows the dynamics of a puller ($\beta=+3$) and its local polymeric environment, characterized by polymer length $N=30$, stiffness $k_b=3000k_BT$ and density $\rho=0.2$, in the reference frame of the moving squirmer.

\twocolumn

\bibliographystyle{eplbib}


\end{document}